# All-analytical semiclassical theory of spaser for plasmonic nanocavity

*Xiao-Lan Zhong* and *Zhi-Yuan Li*[*]

Laboratory of Optical Physics, Institute of Physics, Chinese Academy of Sciences, Beijing 100190, China

[*]*Email address:lizy@aphy.iphy.ac.cn*

**Abstract:** Experimental approaches to manipulate light-matter interaction at nanoscale have quickly advanced in recent years, leading to the demonstration of spaser (surface plasmon amplification by stimulated emission of radiation) in plasmonic nanocavities. Yet, a well-understood analytical theory to better understand and quantitatively explain the connotation of spaser system is urgently needed. Here we develop an all-analytical semiclassical theory to investigate the energy exchange between active materials and fields and the spaser performance in a plasmonic nanocavity. The theory incorporates the four-level atomic rate equations in association with the classical oscillator model for active materials and Maxwell's equations for fields, thus allowing one to uncover the relationship between the characteristics of spaser (the output power, saturation, threshold, etc.) and the nanocavity parameters (quality factor, mode volume, loss, spontaneous emission efficiency, etc.), atomic parameters (number density, linewidth, resonant frequency etc.), and external parameters (pumping rate, etc.). The semiclassical theory has been employed to analyze previous spaser experiments, which shows that a single gold nanoparticle plasmonic nanocavity is very difficult to ignite spaser due to too high threshold. The theory can be commonly used in understanding and designing all novel microlaser, nanolaser, and spaser systems.

## I. Introduction

Recent years have witnessed the rapid development of micro-processing technology, integrated optics, and nanophotonics [1-3]. One of the central issues, the interaction between light and active photonic and plasmonic nanostructured materials has attracted extensive and intensive interest of researches and studies [1-6]. The capability to control light at nanoscale by these active nanostructures has given rise to a rich variety of physical phenomena, such as trapping and manipulation of photons in a resonant nanocavity [1-3], coherent emission, transport, and amplification of surface plasmons [4-6], giant local field enhancement [7,8], compensation of metallic dissipation loss [9,10], and amplification of gain [11,12]. These phenomena can be harnessed for building high-efficiency miniaturized photonic and optoelectronic devices. Through a multi-pronged effort, numerous theoretical and experimental works have been devoted to explore novel ways to miniaturize traditional laser systems and realize nanolasers with tiny footprints and low power consumption. Among them, photonic crystal nancavity lasers [13-18] and plasmonic lasers [19-29] have stood out as two prominent routes toward this fundamental purpose. The former is based on localization and amplification of light within a semiconductor nanocavity with gain media, while the latter is based on so-called surface plasmon amplification by stimulated emission of radiation (spaser) [4,19] in plasmonic nanostructures incorporated with gain media.

In principle, the properties and performances of nanolasers can be understood with the semiclassical physical model of harmonic oscillators coupled to electromagnetic fields. Yet, as the geometries of nanolasers are very complicated involving many subtle nanoscale morphologic features, the electromagnetic fields of laser mode do not have simple spatial profiles, but rather they are far more complex than plane waves or Gaussian beams in traditional laser systems. As a result, it is never an easy thing to describe the interaction of gain media with electromagnetic fields in a simple analytical way as in traditional laser system [30]. Perhaps for this reason, up to now people only largely employ numerical simulation methods, e.g. finite-difference time-domain (FDTD) method, in combination with the atomic rate equations, the dipole approximation model and Maxwell's equations, to investigate nanolasers in several realistic active dielectric and plasmonic systems [31-36]. In contrast, the efforts to build some analytical models to solve the central issues of nanolaser performance are still very rare [37-39]. As is known, a simple, comprehensive, and still quantitatively accurate analytical theory can greatly help to better understand, explain, and predict all concerned important issues of these complicated active nanolaser systems, and then design novel systems of improved performances. In comparison, the route of all numerical simulations will be difficult to extract a clear physics picture about these central issues, although technically accurate data about nanolaser performances can be obtained from the huge consumption of numerical calculations.

In this paper, we report our effort to build up an easy-to-understand all-analytical semiclassical theory for nanolasers by taking into account energy exchange between active materials and fields, power density conservation, and spontaneous and stimulated emission. The theory starts from the basic atomic rate equation in association with the classical oscillator model, considers various aspects of nanocavity parmeters, atomic parameters, and external pumping parameters, and has a final form looking very similar to that for conventional lasers [30]. We will focus on plasmonic lasers where spaser takes place in a plasmonic nanocavity with active materials. The derived all-analytical semiclassical theory can explain the spaser effect in this plasmonic nanolaser system more clearly and precisely.

## II. Theoretical model and analytical solution

We consider a plasmonic nanocavity which is composed of a metallic core, providing for plasmon resonance modes, surrounded by a dielectric shell containing active materials, providing for gain. The active materials are described by general four-level atomic system. The structure, as schematically illustrated in Fig. 1(a), can describe the gain and loss process happening in a general optical nanocavity very well. When the electrons are pumped from the ground-state level with a constant pump rate, spontaneous emission happens immediately. Due to the feedback effect by the surface plasmon resonance of the metallic core, it will cause stimulated emission until a steady-state is reached. The schematic diagram of the atomic system is shown in Fig. 1(b), where level one and two are the lower and upper lasing levels, respectively. Under the dipole approximation, the active materials can be seen as dipoles. It is our aim to develop a methodology to describe the characteristics of light-matter interaction in this nanolaser system. Similar to conventional laser theory [30], we consider atomic transitions in the current nanolaser system quantum mechanically and adopt the model of atomic rate equations to describe these transitions, while handle the radiation of electromagnetic field classically, whose motion follows Maxwell's equations. The interaction between atoms and fields is thus treated semiclassically. Such a semiclassical theory should yield a much more precise description and better prediction of the optical properties of these nanolaser systems than the usual classical theory, where the role of atoms comprising the active materials is described by pheomenological parameter of dielectric permittivity [7,8,11,12].

The occupation numbers of electrons at the atomic levels at each spatial point vary according to the atomic rate equations [29,32,33]:

$$\frac{dN_3}{dt} = W_P N_0 - \frac{N_3}{\tau_{32}}, \tag{1}$$

$$\frac{dN_2}{dt} = \frac{N_3}{\tau_{32}} + \frac{1}{\hbar\omega_a} E_l \cdot \frac{\partial P_{at}}{\partial t} - \frac{N_2}{\tau_{21}}, \tag{2}$$

$$\frac{dN_1}{dt} = \frac{N_2}{\tau_{21}} - \frac{1}{\hbar\omega_a} E_l \cdot \frac{\partial P_{at}}{\partial t} - \frac{N_1}{\tau_{10}}, \tag{3}$$

$$\frac{dN_0}{dt} = \frac{N_1}{\tau_{10}} - W_P N_0 . \tag{4}$$

Eqs. (1) to (4) mean that an external excitation mechanism pumps electrons from the ground-state level ($N_0$) to the third level ($N_3$) at a certain pump rate ($W_P$), which is proportional to the pumping light intensity in the case of optical pumping experiment. After a short lifetime ($\tau_{32}$), electrons transfer nonradiatively into the upper lasing level, i.e. the second level ($N_2$). Electrons can be transferred from the upper to the lower lasing level, i.e. the firs level ($N_1$), by spontaneous and stimulated emission. At last, electrons transfer quickly and nonradiatively from the first level ($N_1$) to the ground-state level ($N_0$). The lifetimes and energies of the upper and lower lasing levels are $\tau_{21}$, $E_2$ and $\tau_{10}$, $E_1$, respectively. The center frequency of the radiation is $\omega_a = (E_2 - E_1)/\hbar$. $E_l$ is the local electric field in the cavity, $P_{at}$ is the electric polarization of atoms, and the term $\frac{1}{\hbar\omega_a} E_l \cdot \frac{\partial P_{at}}{\partial t}$ is the induced radiation rate or excitation rate depending on its sign. As time goes on, the system reaches gradually the steady-state and the steady-state can be described by $dN_i/dt = 0$. The populations at steady-state can be easily solved and written as

$$N_{1,ss} = W_P N_0 \tau_{10} \quad , \tag{5}$$

$$N_{2,ss} = W_P N_0 \tau_{21} - \frac{\omega \tau_{21} \varepsilon_0}{2\hbar\omega_a} \chi_{at}^{"} E_l^2 \quad , \tag{6}$$

$$N_{3,ss} = W_P N_0 \tau_{32} \quad , \tag{7}$$

where $\chi_{at}^{"}$ is the imaginary part of the atomic polarizability.

The population difference between the lower and upper lasing level is

$$\Delta N_{12} = W_P N_0 (\tau_{10} - \tau_{21}) + \frac{\omega \tau_{21} \varepsilon_0}{2\hbar\omega_a} \chi_{at}^{"} E_l^2 . \tag{8}$$

Following the classical harmonic oscillator model, the polarization $P_{at}$ in the presence of an electric field obeys locally the following equation of motion:

$$\frac{d^2 P_{at}(t)}{dt^2} + \Delta\omega_a \frac{dP_{at}(t)}{dt} + \omega_a^2 P_{at}(t) = \Gamma_a \Delta N(t) E_l(t) \quad , \tag{9}$$

where $\Delta\omega_a$ is the linewidth of atomic transition frequency and $\Gamma_a$ is the coupling strength of the polarization to the external electric field. The expression of $\Gamma_a$ is $3\omega_a \varepsilon_h \lambda^3 \gamma_{rad} / 4\pi$, where $\gamma_{rad}$ is the radiative decay rate, $\varepsilon_h$ is the dielectric constant of host material and $\lambda$ is the radiation wavelength.

From Eq. (9) the polarization can be written as

$$P_{at} = \frac{\Gamma_a (N_1 - N_2)}{(\omega_a^2 - \omega^2 + j\omega\Delta\omega_a)} E_l \quad . \tag{10}$$

According to the power density conservation, the storage power density by the upper lasing level $p_{in}$ is equal to the sum of the output power density $p_{out}$, the absorption power density by the cavity $p_{abs}$ and the loss of spontaneous emission power density $p_{loss-spo}$. The output power density and the absorption power density constitute the loss of cavity power density $p_{loss-cav}$. The above relationship can be expressed as

$$p_{in} = p_{out} + p_{abs} + p_{loss-spo} \,, \tag{11}$$

and

$$p_{loss-cav} = p_{out} + p_{abs} \,. \tag{12}$$

Due to the electronic pumping, the power density which can be witnessed and used by the third level can be written as

$$p_{pump} = W_P N_0 \hbar \omega_{30} \,. \tag{13}$$

We introduce a parameter called quantum efficiency $\eta_{qe}$ so that the storage power density by the upper lasing level $p_{in}$ is

$$p_{in} = \eta_{qe} \times p_{pump} \,. \tag{14}$$

As is known, the loss power of the cavity is proportional to $\omega\varepsilon_0 E_l^2 V_{\mathrm{m}} / Q$, where $\omega$ is the resonance frequency of the cavity, $V_{\mathrm{m}}$ is the mode volume and Q is the quality factor (Q-factor) of the cavity. This is the standard definition of the Q-factor of a resonant cavity. We bring in another parameter called cavity loss coupling strength coefficient $\eta_F$, which strongly depends on the geometric and material parameter of the cavity, so that the total loss power of the cavity can be written as

$$P_{loss-cav} = \eta_F \times \frac{\omega \varepsilon_0 E_l^2 V_{\mathrm{m}}}{Q} , \tag{15}$$

and the loss power density is $p_{loss-cav} = P_{loss-cav} / V_c$, where $V_c$ is the cavity volume. It is obvious that a larger value of $\eta_F$ means easier loss of energy power from the cavity.

The spontaneous emission power is only relevant to the population of upper lasing level, and we can get the spontaneous emission power density in the follow expression

$$p_{spo} = \frac{N_{2,ss} \hbar \omega_a}{\tau_{21}} . \tag{16}$$

We notice that the spontaneous emission power is proportional to the upper lasing level population, however, not all of the spontaneous emission power run away from the cavity, most of them are either used to excite stimulated emission or absorbed by the cavity. The escaped spontaneous emission power density from the cavity can be defined as

$$p_{loss-spo} = \eta_{spo} \times p_{spo} , \tag{17}$$

where $\eta_{spo}$ is called the loss spontaneous emission efficiency. Obviously a larger value of $\eta_{spo}$ means stronger loss of spontaneous emission power from the cavity and simultaneously less conversion of this power into the useful laser energy power for the cavity.

Take Eqs. (12), (14), (15) and (17) into consideration and we can get

$$\eta_{qe} \times p_{pump} = \eta_F \times \frac{\omega \varepsilon_0 E_l^2 V_{\mathrm{m}}}{Q V_c} + \eta_{spo} \times p_{spo} . \tag{18}$$

Take the relevant equations into Eq. (18) and then we get

$$\eta_{qe} \times W_P N_0 \hbar\omega_{30} = \eta_F \times \frac{\omega\varepsilon_0 E_l^2 V_{\mathrm{m}}}{QV_c} + \eta_{spo} \times \frac{N_{2,ss}\hbar\omega_a}{\tau_{21}}. \tag{19}$$

We consider the original definition of polarization which is shown below

$$P_{at} = \chi_{at}\varepsilon_0 E_l = (\chi_{at}^{'} + j\chi_{at}^{''})\varepsilon_0 E_l \quad , \tag{20}$$

where $\chi_{at}^{'}$ is the real part of the atomic polarizability.

The above formulae can be combined together to offer solution for various optical properties for the nanolaser system, after some tedious but straightforward algebraic manipulations. We start from the quantity of atomic polarizability $\chi_{at}$. For the sake of simplicity of discussions, we define three parameters:

$$\rho_1 = QV_c\eta_{spo}\omega_a\varepsilon_0\left[\frac{4\omega_a^2\left(\omega_a-\omega\right)^2+\omega^2\Delta\omega_a^2}{\Delta\omega_a}\right], \tag{21a}$$

$$\begin{aligned}\rho_2 = {} & QV_c\Gamma_a W_P N_0\omega\left(\eta_{spo}\omega_a\tau_{10}-\eta_{qe}\omega_{30}\tau_{21}\right) \\ & -2V_{\mathrm{m}}\eta_F\omega_a\varepsilon_0\left[\frac{4\omega_a^2\left(\omega_a-\omega\right)^2+\omega^2\Delta\omega_a^2}{\Delta\omega_a}\right],\end{aligned} \tag{21b}$$

$$\rho_3 = 2\Gamma_a W_P N_0 V_{\mathrm{m}}\eta_F\omega_a\omega\left(\tau_{21}-\tau_{10}\right). \tag{21c}$$

By using of Eqs. (6), (10), (19) and (20), $\chi_{at}^{''}$ can be written as

$$\chi_{at}^{''} = -\frac{\rho_2+\sqrt{\rho_2^2-4\rho_1\rho_3}}{2\rho_1}. \tag{22}$$

Eq. (22) represents the absorbing (or amplifying) part of the atomic response [28]. Respectively, $\chi_{at}^{'}$ and the local electric field can be written as

$$\chi_{at}^{'} = \frac{2\omega_a\left(\omega-\omega_a\right)\chi_{at}^{''}}{\omega\Delta\omega_a}, \tag{23}$$

$$E_l^2 = \frac{\left(\eta_{qe}\times\hbar\omega_{30}-\eta_{spo}\times\hbar\omega_a\right)W_P N_0}{\left(\eta_F\times\omega\varepsilon_0 V_{\mathrm{m}}/QV_c-\eta_{spo}\times\omega\varepsilon_0\chi_{at}^{''}/2\right)}. \tag{24}$$

If the radiation frequency $\omega_a$ is equal to the cavity resonance frequency $\omega$, the expression equations will become much simpler.

Considering the dipole approximation and the tiny nanocavity volume, the absorption power density of the metallic core can be written as

$$p_{abs} = \frac{1}{2}\varepsilon_0 \omega \chi''_{host} E_l^2 . \tag{25}$$

If the loss power density of cavity $p_{loss-cav}$ is larger than the absorption power density of the metallic core $p_{abs}$, which means that the cavity constant $\eta_F$ is larger than $\chi''_{host} V_c Q / 2V_{\mathrm{m}}$, the laser can output from the cavity. We will discuss how to calculate the cavity constant $\eta_F$ later. Now we can deduce the output power density of the nanolaser system, which can be written as

$$p_{out} = \frac{(2\eta_F V_{\mathrm{m}} - Q\chi''_{host})}{2Q} \omega \varepsilon_0 E_l^2 . \tag{26}$$

It is seen that the output power of the active nanocavity is also proportional to the laser field intensity within the cavity. Until now, we have got done the all-analytical semiclassical theory for quantitatively describing the spaser system. The theoretical model has considered all the changeable parameters and it is a generic model enabling us to solve and explain a variety of complex spaser and nanolaser systems. In order to show a clearer physical image, we will give a detailed analysis below.

## III. Analytical and numerical results and discussions

The above equations that comprise the all-analytical semiclassical theory involve lots of parameters, most of which are adjustable. We divide these parameters into three parts: cavity parameters, atomic parameters and external input parameters. As these parameters have encompassed all the geometric and physical details of the nanocavity system, the theory is quite general and can handle various types of nanolasers and various optical problems. The quantum efficiency $\eta_{qe}$, the coupling strength of atomic polarization to the external electric field $\Gamma_a$, the lifetime of each level $\tau_{21}$, $\tau_{10}$ and $\tau_{32}$, and the transition frequency $\omega_a$ and $\omega_{30}$, the linewidth of atomic transition frequency $\Delta\omega_a$ belong to atomic parameters. The cavity loss coupling strength coefficient $\eta_F$, Q-factor Q, cavity volume $V_c$, mode volume $V_{\mathrm{m}}$, the resonance frequency $\omega$, even the loss of spontaneous emission efficiency $\eta_{spo}$ are cavity parameters. The pump rate $W_P$ is the only external input parameter. Besides, there are some constants in this system, e.g. the total population density $N_0$.

In order to obtain clear physical images and insights about the relationship between the spaser properties and the cavity parameters and external excitation parameters, we choose several parameters as reported in reference [22]. The number of dye molecules per nanoparticle is 2700. The core and shell diameters of the plasmonic cavity are 14 nm and 44 nm, respectively, so the cavity volume $V_c$ is $1.436\times10^{-24}\mathrm{m}^3$ and the total population density $N_0$ is $6.255\times10^{25}\mathrm{m}^{-3}$. The radiation frequency $\omega_a$ is $3.5565\times10^{15}Hz$ and the corresponding wavelength is 530 nm, which is close to the resonance wavelength of the cavity. The linewidth of the transition frequency $\Delta\omega_a$ is around $0.04\omega_a$. The lifetime $\tau_{10}$, $\tau_{21}$ and $\tau_{32}$ are chosen as $10^{-9}s$, $10^{-8}s$ and $10^{-9}s$, respectively. The coupling strength of $P_{at}$ to the external electric field $\Gamma_a$ is taken to be $10^{-4}\ \mathrm{C}^2/\mathrm{kg}$ according to references [33]. In our calculations, we find that $\Gamma_a$ does not have a direct influence on the local electric field, the atomic polarizability, and even the output power density. Instead, it is the nature of the atomic system, so we will not discuss the influence of $\Gamma_a$ in this paper.

We first consider the relationship between the local electric field and cavity parameters. The results are shown in Fig. 2 by color contour maps. Each time we only discuss the relationship between two various parameters and the local electric field. Here the pumping rate $W_P$, the Q-factor, the cavity loss coupling strength coefficient $\eta_F$ and the loss spontaneous emission efficiency $\eta_{spo}$ are fixed as $10^4 s^{-1}$, 10, 3 and 6%, respectively. From Fig. 2(a) to (c) we can find that the local electric field increases with the increase of $W_P$ and Q-factor, and with the decrease of $\eta_F$ and $\eta_{spo}$. The relationship between these four parameters and the local electric field can be described by Eq. (24). From Figs. 2(a) and (e) we can find that when $\eta_{spo}$ is large enough, i.e., when most of the spontaneous emission does not participate in the lasing action, the local electric field can achieve saturation phenomenon with the increase of $W_P$ and Q-factor. Otherwise, the saturation phenomenon cannot happen easily. From Fig. 2(b) we can find that when Q-factor is small enough, i.e., when the loss of cavity is large, the local electric field can also achieve saturation phenomenon with $W_P$ increasing. From Figs. 2(c), (d) and (f), the relationship between $\eta_F$ and the local electric field is shown clearly. When $\eta_F$ increases, the local electric field decreases. We should notice one thing here that a larger local electric field does not necessarily mean a higher lasing output power because of the absorption of the cavity, i.e., the absorption of the metallic core.

As is mentioned above, $\chi_{at}^{"}$ means the absorbing (or amplifying) part of the atomic response and it can directly describe the gain and loss of the system. Due to the complex expression of $\chi_{at}^{''}$ by Eq. (22), it is hard to know its quantitative relationship with the cavity parameters directly. With the aid of computer calculation, we obtain some results as shown in Fig. 3. From Figs. 3(a) to (f) we can find that $\chi_{at}^{''}$ increases with the increase of $W_P$, $\eta_F$, $\eta_{spo}$, and with the decrease of Q-factor. The results are quite different from Fig. 2 and this means the absorbing (or amplifying) part of the atomic response does not have a direct relation with the local electric field. From Figs. 3(a) and (c) we can find that the influences of $\eta_F$ and $\eta_{spo}$ on $\chi_{at}^{''}$ are not obvious. The reason is because of the huge influences of pumping rate. We can also find that when $\eta_{spo}$ is large enough, i.e., when most of the spontaneous emission does not participate in the lasing action, $\chi_{at}^{''}$ can achieve the saturation phenomenon with Q-factor increasing. From Figs. 3(c) (d) and (f), the relationship between $\eta_F$ and $\chi_{at}^{''}$ is shown clearly. When $\eta_F$ increases, $\chi_{at}^{''}$ increases. There is an interesting phenomenon here that is different from the long-held general knowledge: A larger local electric field does not necessarily mean a larger gain for the nanocavity.

Next, we discuss an important quantity that one is concerned about very much: The output power density $p_{out}$ of the nanolaser system. This is a characteristic indicator for describing a laser system. The results are shown in Fig. 4. From Figs. 4(a) to (f) we can find that the output power density increases with increasing of pumping rate and $\eta_F$, and with decreasing of Q-factor and $\eta_{spo}$. What is more, from the shadow parts of Figs. 4 (b) and (c) we can find that sometimes no matter how large the pumping rate is, there is still no output power density. It looks like weird at the first glance because as long as the gain is larger than the loss, there should be laser output, and the critical value of the pumping rate corresponds to the threshold. But if we think over it more closely we can find it is not difficult to explain. This spaser system is different from the traditional laser system because the cavity loss comes from the metallic absorption and the loss of spontaneous emission. Whether the metallic absorption or the loss of spontaneous emission, both of them are related to the local electric field, and the local electric field is proportional to the pumping rate. That means with increasing the pumping rate, the loss also increases. The threshold of spaser system comes from the cavity characteristics, i.e. the Q factor and $\eta_F$ instead of large enough pumping rate. From Fig. 4(a) we can find that when $\eta_{spo}$ is large enough, i.e., most of the spontaneous emission does not join in the lasing action, the saturation phenomenon still exists and the output power density can achieve saturation phenomenon with pumping rate increasing. However, from Fig. 4(b), we can also find the similar saturation phenomenon only if Q-factor is less than the threshold. Focusing on the shadow region in Figs. 4 (b) to (f), we can obtain some important conclusions. As is known, the larger Q-factor means the smaller cavity loss and the larger $\eta_F$ corresponds to the easier loss of energy power from the cavity. The function of Q-factor and $\eta_F$ are opposite in influencing the performance of spaser. From Eqs. (11) and (12) we can find that $p_{out} = p_{loss-cav} - p_{abs}$. As is mentioned above, both of the metallic absorption and the loss of spontaneous emission are related to the local electric field intensity, either directly or indirectly. Therefore, this spaser system shows threshold related with cavity parameters, i.e., Q-factor and $\eta_F$.

In order to have a clarified physical image to illustrate the nature of the plasmonic nanocavity, we calculate the relationship between power density, Q-factor and $\eta_F$. The results are shown in Fig. 5. From Figs. 5(a) to (d) the power densities are $p_{loss-cav}$, $p_{abs}$, $p_{loss-spo}$ and the needed threshold power density of the plasmonic nanocavity $p_{\mathrm{thr}}$, where $p_{\mathrm{thr}}=p_{abs}+p_{loss-spo}$. From Eqs. (15) and (25), we can find that both $p_{loss-cav}$ and $p_{abs}$ are proportional to the local electric field intensity. The calculation results show the same phenomena in Figs. 5(a) and (b). From Fig. 5(c) we can find that $p_{loss-spo}$ increases with increasing $\eta_F$ and with decreasing Q-factor. From Eq. (11) we can find when $p_{in}$ is larger than $p_{\mathrm{thr}}$, the laser can output from the spaser system. In our model, $p_{in}$ is equal to $2.3458\times10^{11}\ \mathrm{W/m^3}$, which is marked in Fig. 5(d). Comparing with Fig. 4(c), we can find when $p_{\mathrm{thr}}$ is larger than $2.3458\times10^{11}\ \mathrm{W/m^3}$, the output power density is zero.

Another famous phenomenon which is often concerned about is the population difference saturation. Using the all-analytical semiclassical theory we can get the relationship between population saturation and pumping rate. The result is shown in Fig. 6. Here the Q-factor, $\eta_F$ and $\eta_{spo}$ are fixed as 10, 3 and 6%, respectively. From Fig. 6 we can find that when the pumping rate increases, the population difference gradually becomes larger and larger until saturation phenomenon happens.

## IV. Application to analysis of practical experiments

To further illustrate the power of the all-analytical semiclassical theory in handling practical problems of nanolasers, we consider M. A. Noginov's experiments [22], which present the first demonstration of spaser. The spaser system is a plasmonic core-shell nanoparticle, which is composed of a gold core (diameter 14 nm), providing for plasmon modes, surrounded by a silica shell (thickness 15 nm) containing the organic dye Oregon Green 488 (OG-488) (density $6.25\times10^{19}\,\mathrm{cm^{-3}}$), providing for gain. In the experiment, the spectral and temporal characteristics of light leaking from the particles suspended in solution when optically pumped by nanosecond laser were measured. The narrowing of radiation linewidth and linear increase of the magnitude of the resonant peak were observed and attributed to the ignition of spaser from a single plasmonic nanoparticle instead from a collective group of nanoparticles [22]. However, this point has raised controversy and it is our target of theoretical evaluation by using the all-analytical semiclassical theory. To solve this problem theoretically, we adopt a model as schematically depicted in Fig. 1, where the geometric parameters of the nanoparticle are explicitly shown.

To quantitatively handle this spaser problem, we first employ finite-difference time-domain (FDTD) method to calculate various cavity parameters. Considering the dipole approximation method, these OG-488 four-level atoms can be seen as dipoles which distribute uniformly around the metallic particle. Through the FDTD calculation, we can obtain the single-dipole emission power $P_0$. The single-dipole emission power is an external parameter related with the pumping rate. The initial power density then can be written as $p_{in} = N_0 P_0$ and the pumping rate can be written as $W_P = P_0 / \eta_{qe} \times \hbar\omega_{30}$. Taking into account the atomic parameters, we can obtain $P_0 \approx 1.46\times10^{-15}\,\mathrm{W}$ and $W_P \approx 4\times10^{3}\,\mathrm{s^{-1}}$. Next we determine the loss spontaneous emission efficiency $\eta_{spo}$. Considering the different positions and dipole polarization angles, we take average of the radiation power and calculate the value of $\eta_{spo}$. The dipole radiation power $P_{rad}(\theta, r)$ shows resonance at $\theta = 90°$, where the loss spontaneous emission of the spaser system becomes negligible. Considering different positions and angles, we take average of the radiation power and calculate the loss spontaneous emission efficiency, which is $\eta_{spo} \approx 6\%$. The Q-factor of this plasmonic cavity structure is about 10 and the modal volume is about $4.831\times10^{-6}\,\mu\text{m}^3$. The absorption power $P_{abs0}$ by the metallic particle can also be calculated by FDTD method with single-dipole source. Considering different positions and angles, we also need to take average of the absorption power. We can get $\overline{P_{abs0}} \approx 9.448\times10^{-14}\,\text{W}$ and the total absorption power density $p_{abs-fdtd} = N_0\overline{P_{abs0}} \approx 5.91\times10^{12}\ \text{W/m}^3$.

From now on, the only unknown value is the cavity loss coupling strength coefficient $\eta_F$. From the above complete semiclassical theory, we can get the relationship between absorption power density $p_{abs-theory}$ and $\eta_F$ when other quantities are fixed. From the FDTD method we can get the determined total absorption power density $p_{abs-fdtd}$ which is shown above. For the same system, no matter we use which method, either all-analytical semiclassical theory or all-numerical simulation method, the total absorption power density is determined. Comparing the numerical results with the all-analytical theoretical results, we can always find $p_{abs-theory} = p_{abs-fdtd}$ which corresponds with the only determined $\eta_F$. Through calculation we can determine $\eta_F$=0.04 in this system. Recalling Fig. 4(d), we can find that in that case, there is no output power density of spaser from this plasmonic system. We comment that the measurement results in Ref. [22] are more likely not related with the spaser for a single plasmonic nanocavit. Instead, the observed laser performance (sharply narrowing of the spectral response of emission light) might be attributed to other factors, such as random lasers due to the collective action of a group of metal nanoparticles. Our all-analytical semiclassical theory strongly suggests that in order to observe spaser in this single plasmonic nanocavity, the atomic density of gain materials must be increased by two orders of magnitude, so that a sufficiently large gain could be achieved.

## V. Conclusions

We have investigated the interaction between light and four-level atomic system embedded within a metallic particle and the spaser properties of this active plasmonic nanocavity system. By solving the coupled equations encompassing the atomic rate equation, the classical oscillator model, and Maxwell's equations, and introducing several parameters that merely depend on the geometric and physical properties of the nanocavity, we have constructed an all-analytical semiclassical theory for describing the energy exchange between active materials and fields, light emission, and spaser performance in the plasmonic nanocavity. The theory incorporates the atomic rate equation in association with the classical oscillator model for active materials and Maxwell's equations for fields, thus allowing one to uncover the relationship between the characteristics of spaser (the output power, saturation, threshold, etc.) and the nanocavity parameters (quality factor, mode volume, loss, spontaneous emission efficiency, etc.), atomic parameters (number density, linewidth, resonant frequency etc.), and external parameters (pumping rate, etc.). As a result, the all-analytical semiclassical theory can handle various nanocavity systems (semiconductor nanocavity, plasmonic nanocavity, semiconductor nanowires) consisting of various active materias (atoms, molecules, ions, and semiconductors). Through detailed calculation and analysis, several remarkable things about the spaser performance are discovered. The spaser system has the all characteristics of the traditional laser system, e.g. the saturation phenomenon and the threshold. The semiclassical theory has been employed to analyze previous spaser experiments, which shows that a single gold nanoparticle plasmonic nanocavity is very difficult to ignite spaser due to too high threshold. As the all-analytical semiclassical theory has a simple formalism that looks like the conventional laser theory, it can offer an easy-to-understand yet sufficiently accurate means to understand and explain the behavior of spaser in plasmonic nanocavities, and will be very useful in designing novel spaser devices with high performance. Furthermore, as this universal theory has involved a lot of model parameters, it is expected to be applicable to many different microlaser, nanolaser, and spaser systems.

**Acknowledgements:**

This work is supported by the 973 Program of China at Nos. 2011CB922002 and 2013CB632704, the Knowledge Innovation Program of the Chinese Academy of Sciences (No. Y1V2013L11).

**References:**

1. S. Noda, A. Chutinan, and M. Imada, "Trapping and emission of photons by a single defect in a photonic bandgap structure", Nature **407**, 608-610 (2000).

2. K. J. Vahala, "Optical microcavities", Nature **424**, 839-846 (2003).

3. Zhi-Yuan Li, "Nanophotonics in China: Overviews and Highlights", Frontiers of Physics **7**, 601-631 (2012).

4. D. J. Bergman and M. I. Stockman, "Surface plasmon amplification by stimulated emission of radiation: Quantum generation of coherent surface plasmons in nanosystems", Phys. Rev. Lett. **90**, 027402 (2003).

5. A. V. Akimov, A. Mukherjee, C. L. Yu, D. E. Chang, A. S. Zibrov, P. R. Hemmer, H. Park, and M. D. Lukin, "Generation of single optical plasmons in metallic nanowires coupled to quantum dots", Nature **450**, 402 (2007).

6. M. I. Stockman, "Nanoplasmonics: past, present, and glimpse into future", Opt. Express **19**, 22029-22106 (2011).

7. Z. Y. Li and Y. Xia, "Metal nanoparticles with gain toward single-molecule detection by surface-enhanced raman scattering", Nano Lett. **10**, 243-249 (2010).

8. S. Y. Liu, J. Li, F. Zhou, L. Gan and Z. Y. Li, "Efficient surface plasmon amplification from gain-assisted gold nanorods", Opt. Lett. **36**, 1296-1298 (2011).

9. Yu-Hui Chen, Jia-Fang Li, Ming-Liang Ren, Ben-Li Wang, Jin-Xin Fu, Si-Yun Liu, and Zhi-Yuan Li, "Direct observation of amplified spontaneous emission of surface plasmon polaritons at metal/dielectric interfaces", Appl. Phys. Lett. **98**, 261912 (2011).

10. B. Peng, Q. Zhang, X. Liu, Y. Ji, H. V. Demir, C. H. A. Huan, T. C. Sum and Q. Xiong, "Fluorophore-doped core multishell spherical plasmonic nanocavities: resonant energy transfer toward a loss compensation", ACS Nano **6**, 6250–6259 (2012).

11. D. B. Li and C. Z. Ning, "Giant modal gain, amplified surface plasmon-polariton propagation, and slowing down of energy velocity in a metal-semiconductor-metal structure", Phys. Rev. B **80**, 153304 (2009)

12. X. L. Zhong and Z. Y. Li, "Plasmon enhanced light amplification in metal–insulator–metal waveguides with gain," J. Opt. **14**, 055002 (2012).

13. O. Painter, R. K. Lee, A. Scherer, J. D. O'Brien, P. D. Dapkus, and I. Kim, "Two-dimensional photonics band-gap defect mode laser", Science **284**, 1819-1821 (1999).

14. H. G. Park, S. H. Kim, S. H. Kwon, Y. G. Ju, J. K. Yang, J. H. Baek, S. B. Kim, and Y. H. Lee, "Electrically driven single-cell photonic crystal lasers", Science **305**, 1444-1447 (2004).

15. H. Altug, D. Englund, and J. Vuckovic, "Ultrafast photonic crystal nanocavity laser", Nature Phys. 2, 484–488 (2006).

16. S. Matsuo, A. Shinya, T. Kakitsuka, K. Nozaki, T. Segawa, T. Sato, Y. Kawaguchi, and M. Notomi, "High-speed ultracompact buried heterostructure photonic-crystal laser with 13 fJ of energy consumed per bit transmitted", Nature Photonics 4, 648-654 (2010).

17. B. Ellis, M. A. Mayer, G. Shambat, T. Sarmiento, J. Harris, E. E. Haller, and J. Vuckovic, "Ultralow-threshold electrically pumped quantum-dot photonic crystal nanocavity laser", Nature Photonics **5**, 297-300 (2011).

18. A. Tandaechanurat, S. Ishida, D. Guimard, M. Nomura, S. Iwamoto, and Y. Arakawa, "Lasing oscillation in a three-dimensional photonic crystal nanocavity with a complete bandgap", Nature Photonics **5**, 91-94 (2011).

19. N. I. Zheludev, S. L. Prosvirnin, N. Papasimakis and V. A. Fedotov, "Lasing spaser", Nature Photonics **2**, 351-354 (2008).

20. P. Berini and I. D. Leon, "Surface plasmon–polariton amplifiers and lasers", Nature Photonics **6**, 16-24 (2012).

21. K. Ding and C. Z. Ning, "Metallic subwavelength-cavity semiconductor nanolasers", Light: Science & Applications **1**, e20 (2012).

22. M. A. Noginov, G. Zhu, A. M. Belgrave, R. Bakker, V. M. Shalaev, E. E. Narimanov, S. Stout, E. Herz,T. Suteewong and U. Wiesner, " Demonstration of a spaser-based nanolaser", Nature **460**,1110-1113 (2009).

23. R. F. Oulton, V. J. Sorger, T. Zentgraf, R. M. Ma, C. Gladden, L. Dai, G. Bartal and X. Zhang, "Plasmon lasers at deep subwavelength scale", Nature **461**, 629-632 (2009).

24. M. T. Hill, M. Marell, E. S. P. Leong, B. Smalbrugge, Y. C. Zhu, M. H. Sun, P. J. van Veldhoven, E. J. Geluk, F. Karouta, Y. S. Oei, R. Notzel, C. Z. Ning, and M. K. Smit, "Lasing in metal-insulator-metal sub-wavelength plasmonic

Waveguides", Opt. Express **17,** 11107–11112 (2009).

25. M. P. Nezhad, A. Simic, O. Bondarenko, B. Slutsky, A. Mizrahi, L. Feng, V. Lomakin, and Y. Fainman, ":Room-temperature subwavelength metallo-dielectric lasers", Nature Photonics **4**, 395–399 (2010).

26. J. H. Lee, M. Khajavikhan, A. Simic, Q. Gu, O. Bondarenko, B. Slutsky, M. P. Nezhad, and Y. Fainman, "Electrically pumped sub-wavelength metallo-dielectric pedestal pillar lasers", Opt Express **19**, 21524–21531 (2009).

27. K. Yu, A. Lakhani, and M. C. Wu, "Subwavelength metal-optic semiconductor nanopatch lasers", Opt Express **18**, 8790–8799 (2010).

28. R. M. Ma, R. F. Oulton, V. J. Sorger, G. Bartal, and X. Zhang, "Room temperature sub-diffraction-limited plasmon laser by total internal reflection", Nature Mater. **10,** 110–113 (2011).

29. J. B. Khurgin and G. Sun, "Injection pumped single mode surface plasmon generators: threshold, linewidth, and coherence", Opt. Express **20**, 15309-15325 (2012).

30. A. E. Siegman, Lasers (Hill Valley, California, 1986), Chaps. 2, 3, 6, and 13

31. X. Jiang and C. M. Soukoulis, "Time dependent theory for random lasers", Phys. Rev. Lett. **85**, 70-73 (2000).

32. P. Bermel, E. Lidorikis, Y. Fink and J. D. Joannopoulos, "Active materials embedded in photonic crystals and coupled to electromagnetic radiation", Phys. Rev. B **73**, 165125 (2006).

33. A. Fang, Th. Koschny and C. M. Soukoulis, "Self-consistent calculations of loss-compensated fishnet metamaterials", Phys. Rev. B **82**, 121102(R) (2010).

34. S. Wuestner, A. Pusch, K. L. Tsakmakidis, J. M. Hamm, and O. Hess, "Overcoming losses with gain in a negative refractive index metamaterial", Phys. Rev. Lett. **105**, 127401 (2010).

35. J. M. Hamm, S. Wuestner, K. L. Tsakmakidis, and O. Hess, "Theory of light amplification in active fishnet metamaterials", Phys. Rev. Lett. **107**, 167405 (2011).

36. Z. Huang, Th. Koschny and C. M. Soukoulis, "Theory of pump-probe experiments of metallic metamaterials coupled to a gain medium", Phys. Rev. Lett. **108**, 187402 (2012).

37. H. E. Türeci, A. D. Stone, and B. Collier, “Self-consistent multimode lasing theory for complex or random lasing media”, Phys. Rev. A **74**, 043822 (2006).

38. M. I. Stockman, “The spaser as a nanoscale quantum generator and ultrafast amplifier”, J. Opt. **12**, 024004 (2010).

39. A. Cerjan, Y. Chong, L. Ge, and A. D. Stone, “Steady-state ab initio laser theory for N-level lasers”, Optics Express **20**, 474-488 (2011).

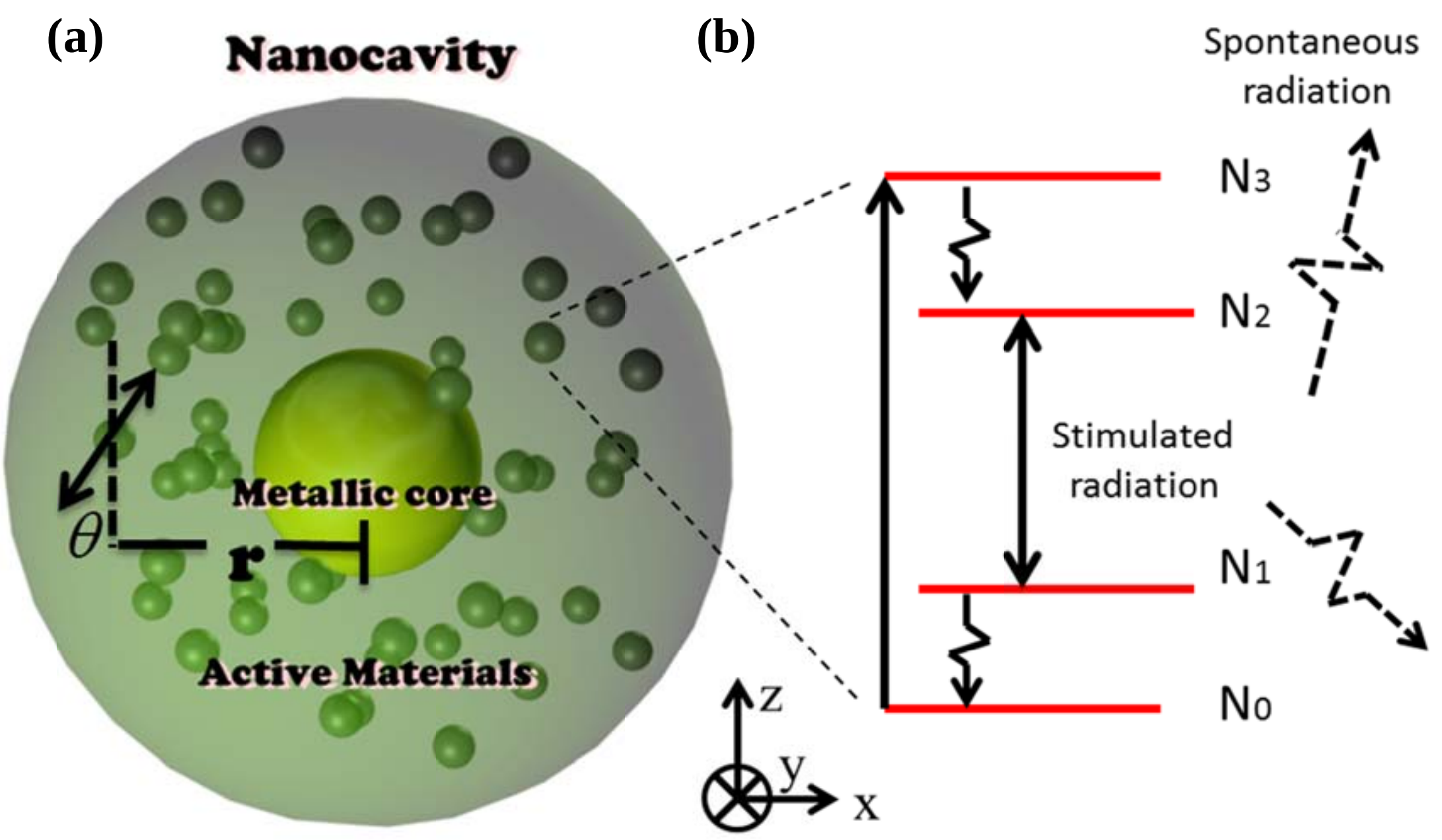

Fig. 1 Schematic plot of the light-matter interaction of a general four-level atomic system with a metallic nanocavity. (a) Geometry of the plasmonic nanocavity, which is a core-shell nanoparticle consisting of a metallic core and dielectric shell doped with four-level atoms as active materials. Each atom is modeled as a radiation dipole. $\theta$ is the angle of the dipole polarization and $r$ is the distance between the dipole and the center of the metallic core; (b) Schematic of a general four-level atomic system which describe the spontaneous and stimulated radiation.

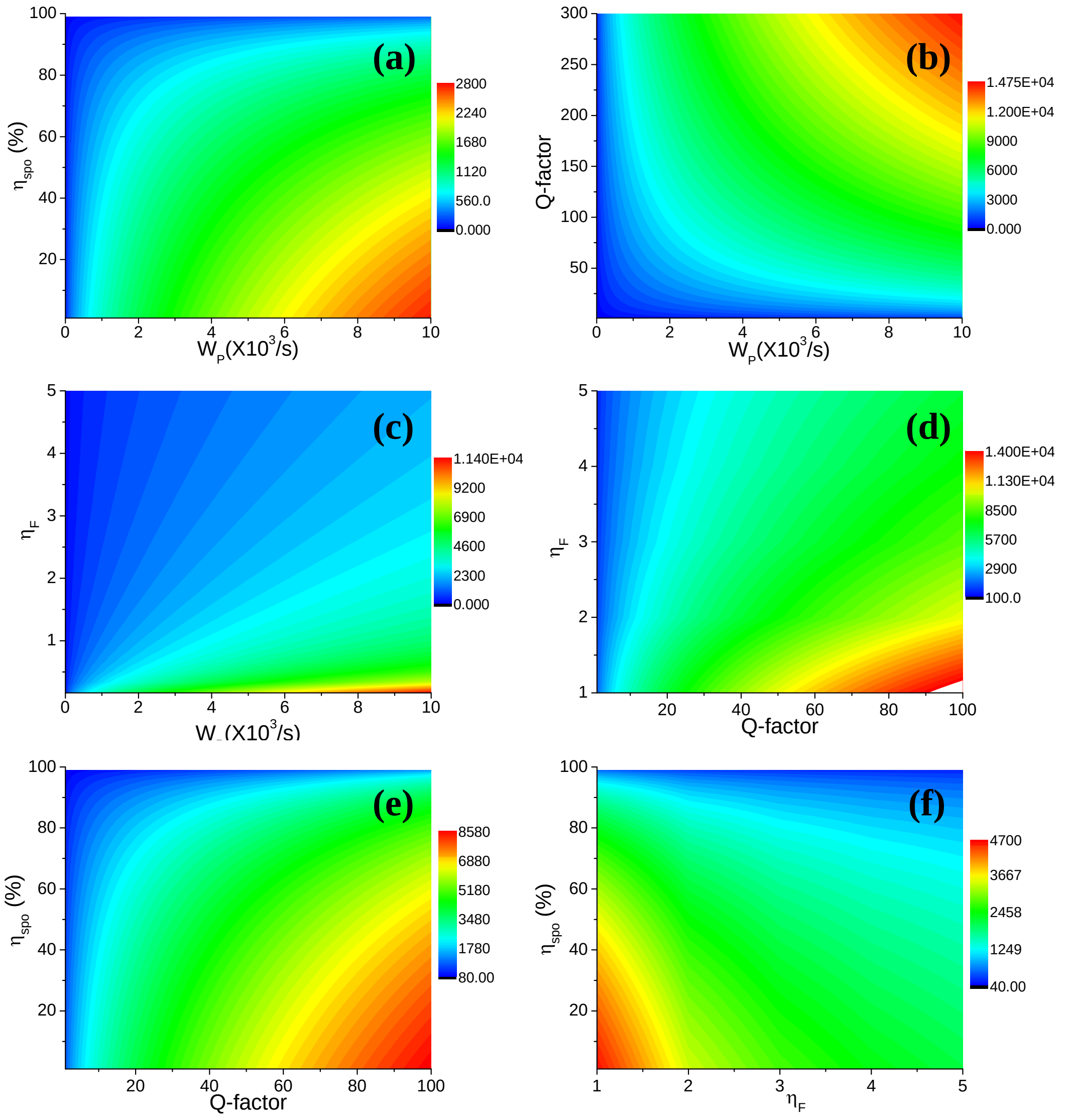

Fig. 2 Contour maps of the local electric field amplitude $E_l$ (in unit of V/m) as functions of various cavity parameters. (a) $E_l$ as functions of pumping rate $W_p$ and $\eta_{spo}$; (b) $E_l$ as functions of $W_p$ and Q-factor; (c) $E_l$ as functions of $W_p$ and $\eta_F$; (d) $E_l$ as functions of Q-factor and $\eta_F$; (e) $E_l$ as functions of Q-factor and $\eta_{spo}$; (f) $E_l$ as functions of $\eta_F$ and $\eta_{spo}$.

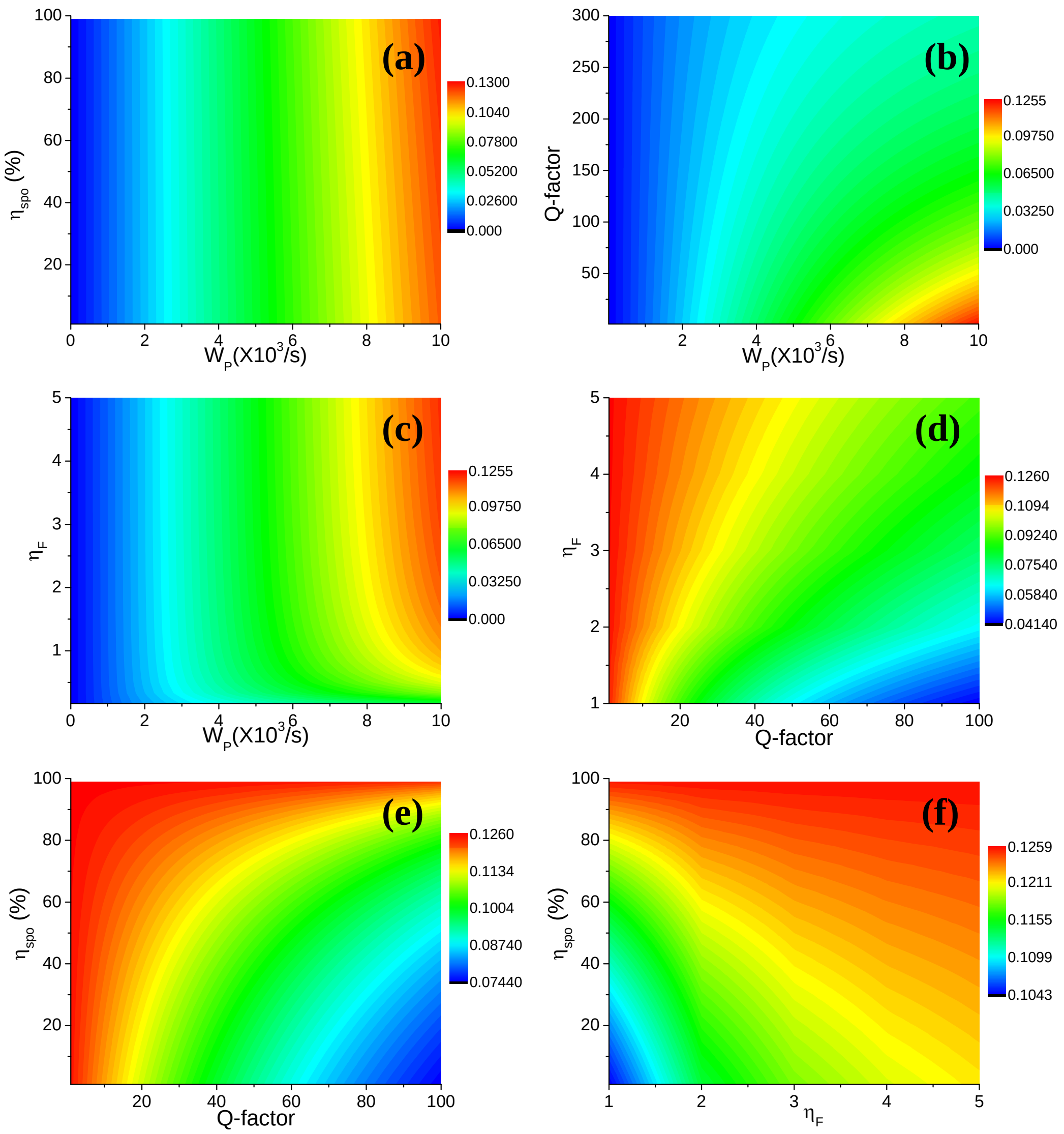

Fig. 3 Contour maps of the imaginary part of atomic polarizability $\chi_{at}''$ as functions of various cavity parameters. (a) $\chi_{at}''$ as functions of pumping rate $W_p$ and $\eta_{spo}$; (b) $\chi_{at}''$ as functions of $W_p$ and Q-factor; (c) $\chi_{at}''$ as functions of $W_p$ and $\eta_F$; (d) $\chi_{at}''$ as functions of Q-factor and $\eta_F$; (e) $\chi_{at}''$ as functions of Q-factor and $\eta_{spo}$; (f) $\chi_{at}''$ as functions of $\eta_F$ and $\eta_{spo}$.

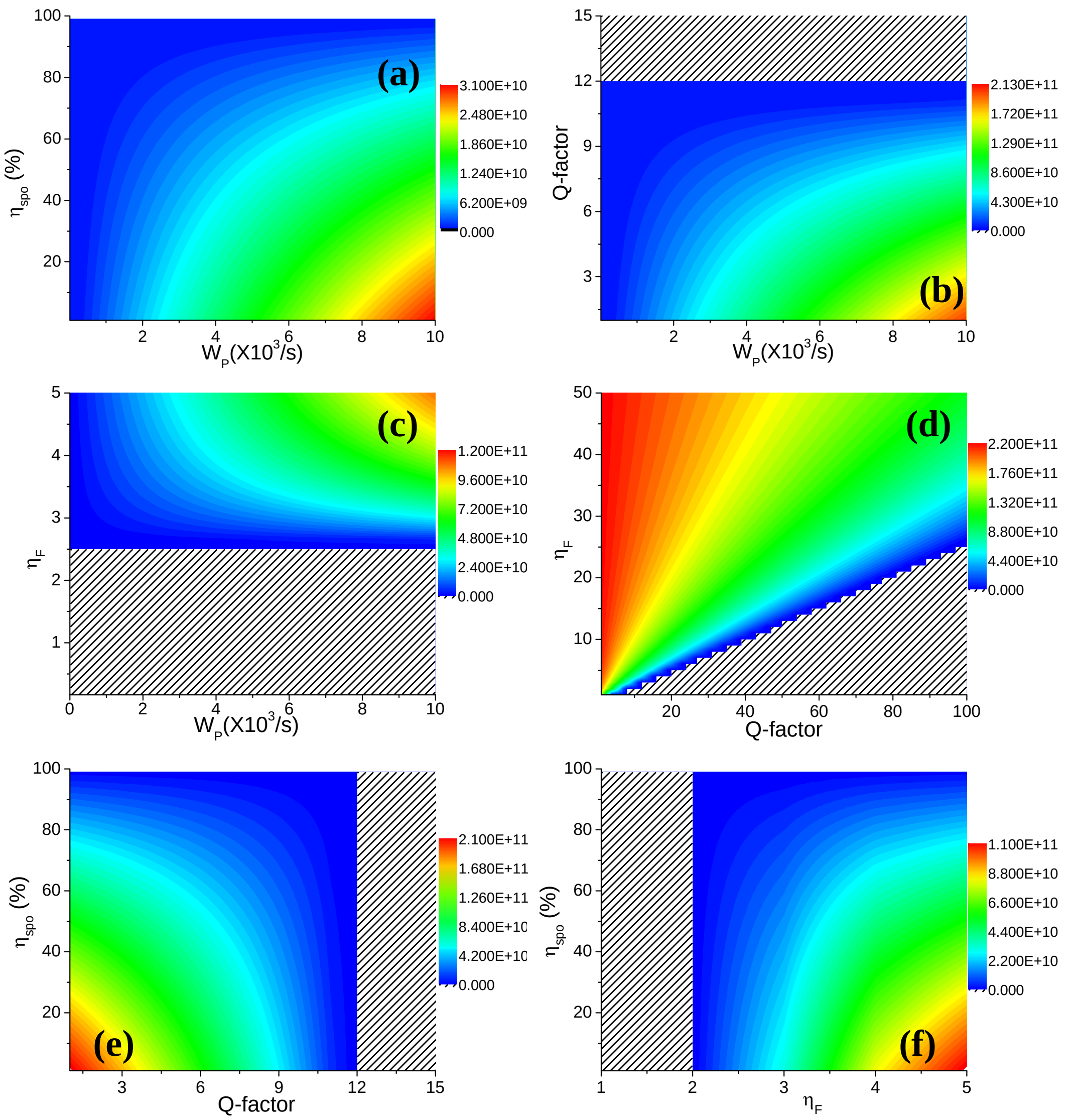

Fig. 4 Contour maps of the laser output power density $p_{out}$ (in unit of W/m$^3$) as functions of various cavity parameters. (a) $p_{out}$ as functions of the pumping rate $W_p$ and $\eta_{spo}$; (b) $p_{out}$ as functions of $W_p$ and Q-factor; (c) $p_{out}$ as functions of $W_p$ and $\eta_F$; (d) $p_{out}$ as functions of Q-factor and $\eta_F$; (e) $p_{out}$ as functions of Q-factor and $\eta_{spo}$; (f) $p_{out}$ as functions of $\eta_F$ and $\eta_{spo}$. The shadow regions mean that the output power density is zero.

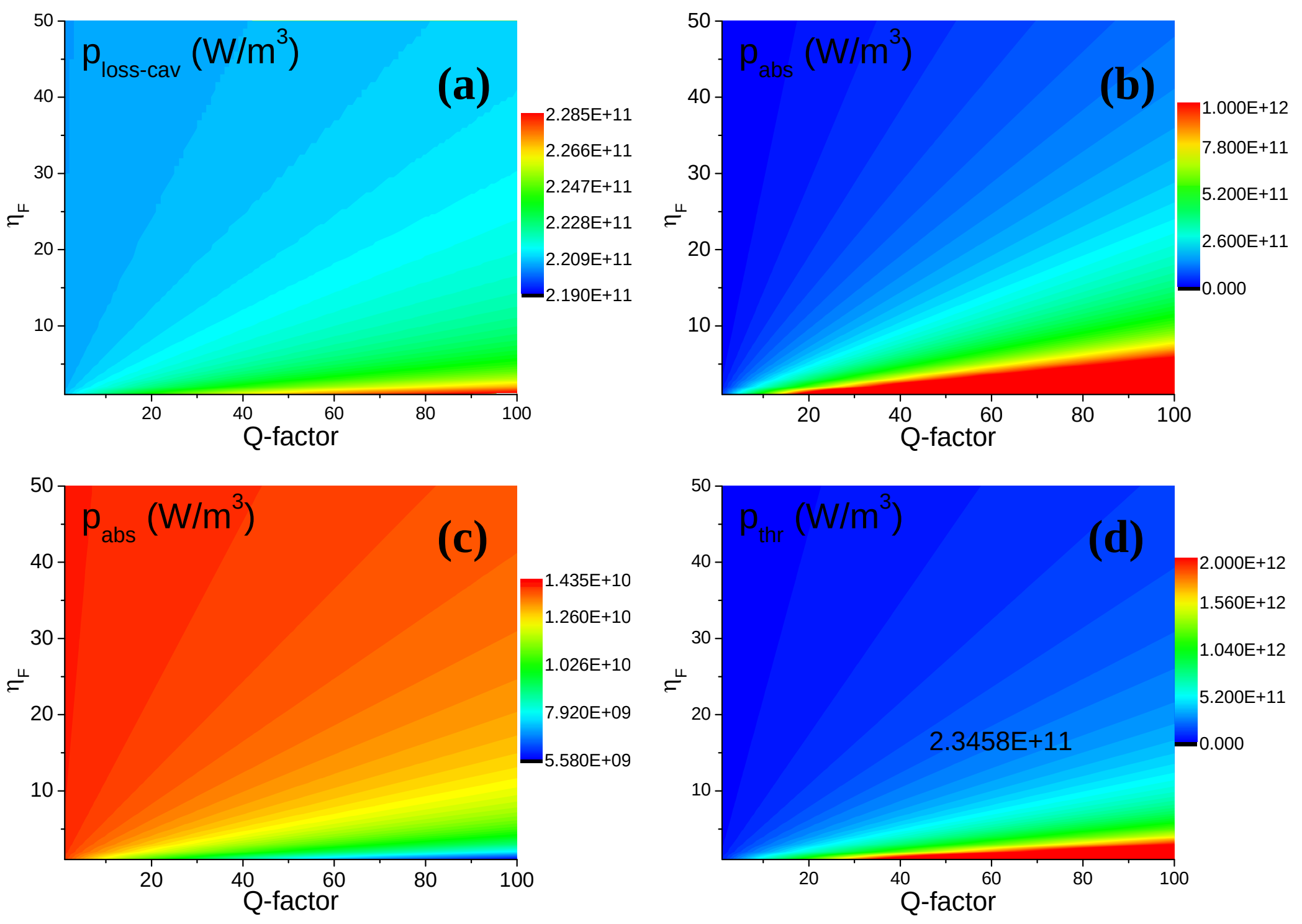

Fig. 5 Contour plot of various power densities of the active plasmonic nanocavity as functions of Q-factor and $\eta_F$. (a) The loss of cavity power density $p_{loss-cav}$; (b) The absorption power density of the metallic core $p_{abs}$; (c) The loss of spontaneous emission power density $p_{loss-spo}$; and (d) The needed threshold power density of the nanocavity $p_{\mathrm{thr}}$.

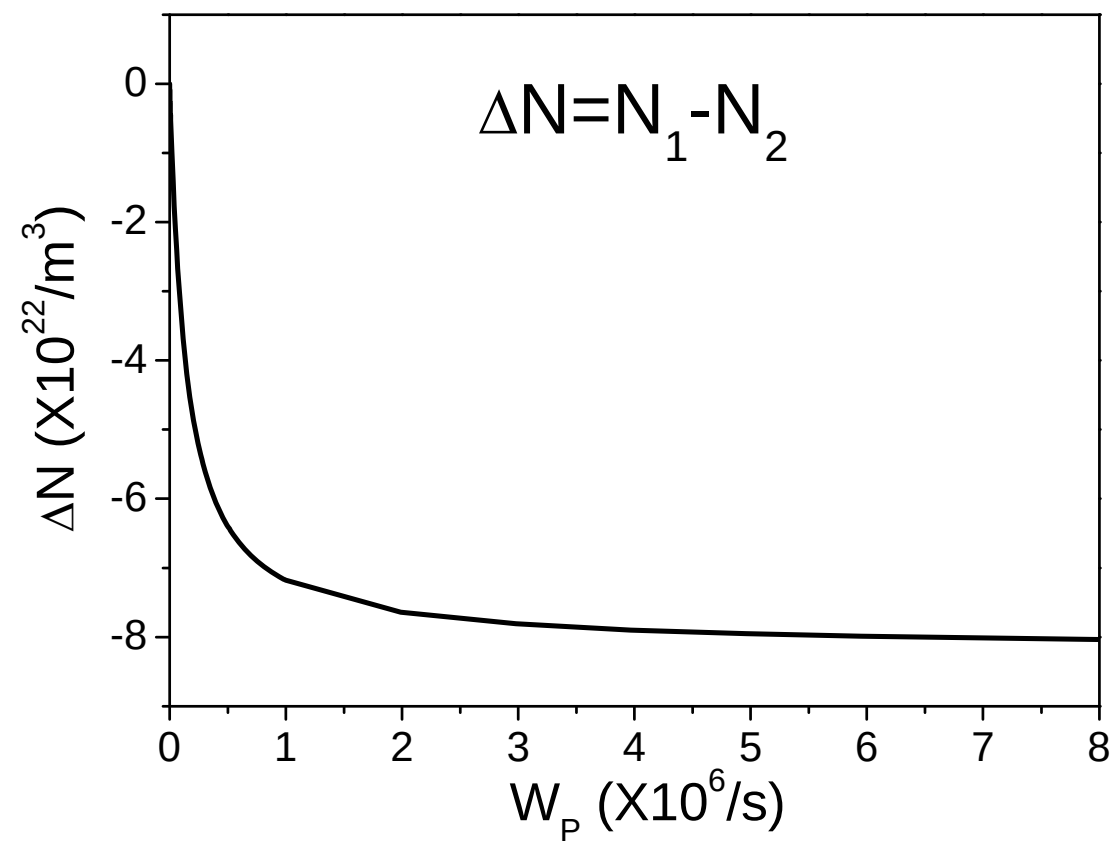

Fig. 6 The relationship between the population difference and pumping rate.